\documentclass[prb,groupedaddress,10pt,twocolumn,reprint,amsmath,amssymb,floatfix]{revtex4-1}

\usepackage{graphicx}
\usepackage{dcolumn}
\usepackage{bm}
\usepackage[usenames, dvipsnames]{color}
\usepackage[utf8]{inputenc}
\usepackage{lettrine}
\usepackage{hyperref}
\usepackage[mathlines]{lineno}

\begin{document}

\title{Sound of Interfacial Flows: Unraveling the Forces Shaping Fast Capillary Flows using their Acoustic Signature}

\author{Adrien Bussonnière$^{1, 2}$}
\author{Arnaud Antkowiak$^{1}$}
\author{François Ollivier$^{1}$}
\author{Michaël Baudoin$^{2}$}
\author{Régis Wunenburger$^{1}$}
\affiliation{$^1$Sorbonne Université, CNRS, Institut Jean le Rond $\partial$’Alembert, UMR 7190, F-75005 Paris, France\\
$^2$Univ. Lille, CNRS, ECLille, ISEN, Univ. Valenciennes, UMR 8520 - IEMN, F-59000 Lille, France}

\date{\today}
\maketitle

\textbf{
Many familiar events feature a distinctive sound: paper crumpling\cite{Houle1996} or tearing\cite{Salminen2002}, squeaking doors\cite{Rubinstein2004}, drumming rain\cite{Franz1959,Gordillo2018} or boiling water\cite{Aljishi1991}.
Such characteristic sounds actually carry a profusion of informations about the fleeting physical processes
at the root of acoustic emission, which appears appealing especially in situations precluding direct or in-situ measurements, such as e.g. the rupture of micron-thick liquid sheet.
Here we report on such a link between fast interfacial hydrodynamics and sound. 
The acoustic emission of a bursting soap bubble is captured by means of antennae and deciphered with the conceptual framework of aeroacoustics. This reveals that capillary forces, thin-film hydrodynamics, but also out-of-equilibrium surfactants dynamics all shape the capillary burst sound. 
Whereas ultra-fast imagery only captures the shapes of flows, the acoustic signature radiated by hydrodynamical forces offers a timely complement for it allows a direct experimental access to these dynamical quantities.
}
\lettrine{F}{orms} and forces in flows with interfaces are intertwined, notably
because interfaces, more than just geometrical boundaries, exert forces. As
Rayleigh noted in 1891, a trademark of these flows is that they ``pass so quickly
so as to elude ordinary means of observation''\cite{Rayleigh1891}. This represents
of course a major hurdle in the experimental characterization of these interfacial
flows, and this probably explains why their understanding has been so strongly
linked to the development of imaging techniques, dating back to the rolling
stroboscope of Savart which allowed to see the disintegration of a liquid stream
into droplets\cite{Savart1833}, the arising of spark photography at the end of the
nineteenth century which revealed the instantaneous rupture of an underwater gas
stream into bubbles\cite{Rayleigh1891} or the shape of a
splash\cite{Worthington1908}, the first ultra-fast camera that monitored the
bursting of a soap bubble at 2,000 frames per second in 1907, tinkered with a
rolling cardboard and repeated sparks in Etienne-Jules Marey
workshop\cite{Lefebvre1995} up to the development of modern ultra-fast imaging used
since the fifties to capture splashes\cite{Engel1955} and soap film
rupture\cite{Ranz1959} and continuously developed since then (see
\citet{Thoroddsen2008} for a review). If forms only provide a partial
picture of these flows, it appears that a direct experimental access to the forces
with e.g. the immersion of pressure sensors in thin rupturing films seems
impracticable. Conversely, time-varying or moving forces radiate acoustic waves.
Here we show on a model bursting soap bubble experiment that these acoustic waves
are detectable and that their analysis allows to find the locus of rupture of the
bubble, and to give time-resolved informations
about  the thickness profile of the bubble, the inner thin-film hydrodynamics and dynamical change in surface tension due
to surfactants reorganization across the bubble during bursting.

\begin{figure}[hbt]
\centering\includegraphics[width=88mm]{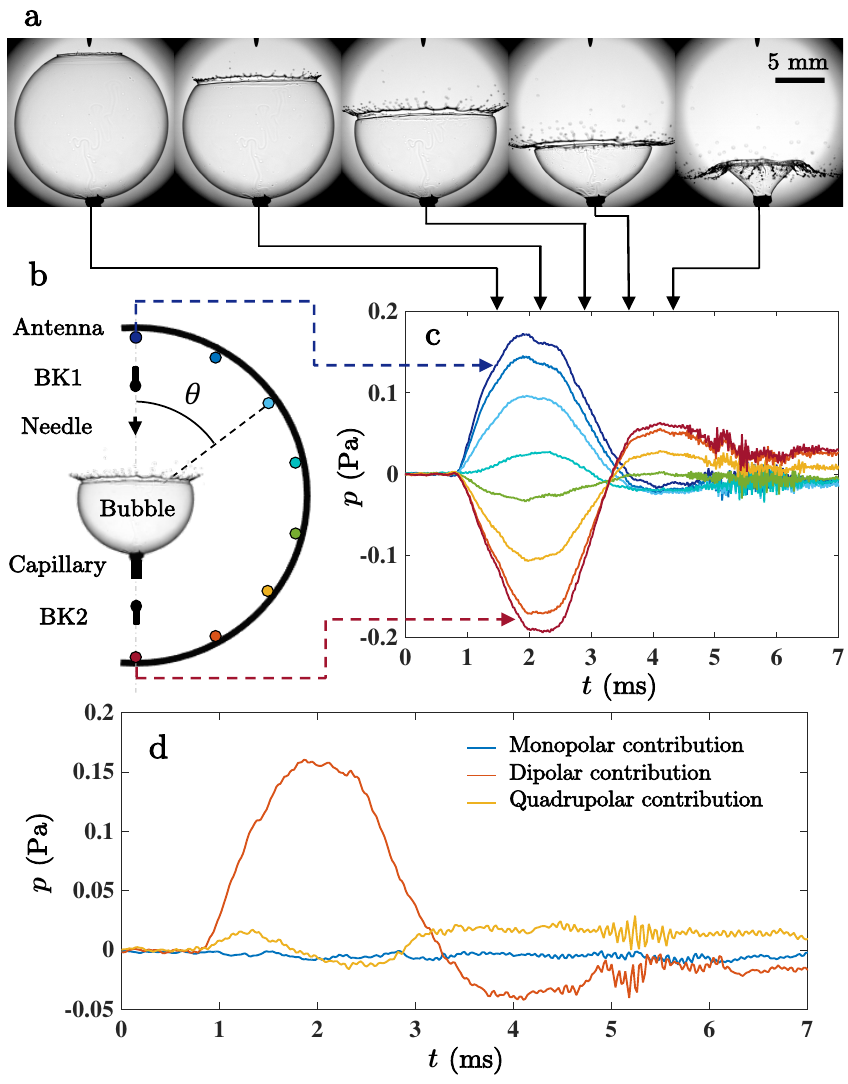}
\caption{\textbf{Directivity and dipolar nature of the acoustic emission. a}, High speed images of the bursting of a fresh 2~mL soap bubble triggered using a needle visible on top of the pictures. \textbf{b}, sketch of the setup showing one of the three circular acoustic antennae each made of eight MEMS microphones (coloured circles) used to measure the sound directivity and the two large bandwidth microphones labelled BK1 and BK2. \textbf{c}, the eight pressure signals acquired by the 44~mm radius acoustic antenna during bubble bursting. The acquisition times of the bubble pictures are indicated by black arrows. \textbf{d}, monopolar, dipolar and quadrupolar contributions to the pressure signal measured by the MEMS microphone located on the bubble top, computed from the 24 signals acquired using the acoustic antennae (see Sup. Mat.).}
\label{fig1}
\end{figure}

\begin{figure}[ht]
\centering\includegraphics[width=88mm]{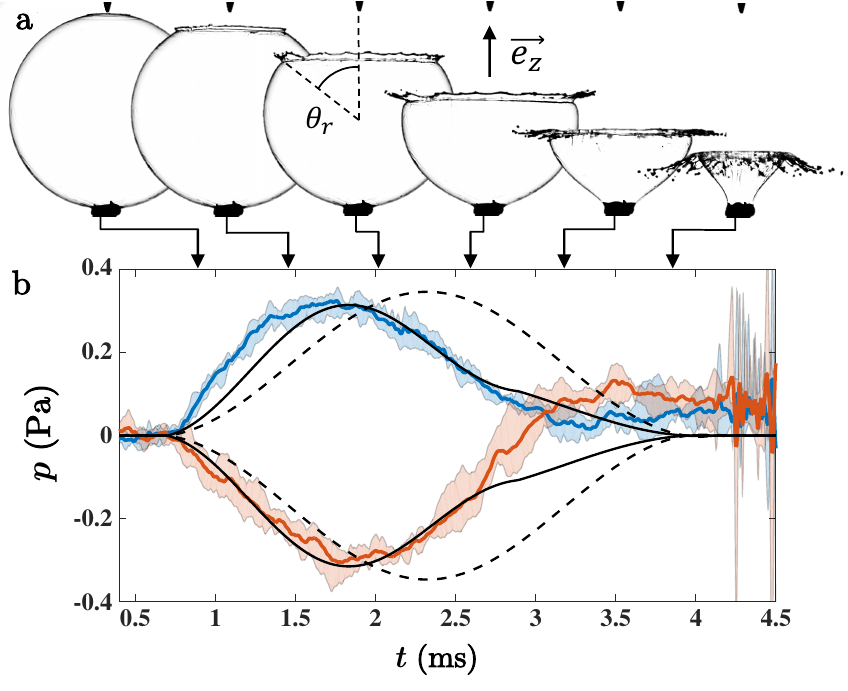}
\caption{\textbf{Confrontation with models of acoustic emission in the case of fresh bubbles. a,} high-speed images of the bursting of a 1~mL, fresh bubble triggered by a needle. \textbf{b,} pressure signals acquired during the bursting of a 1~mL, fresh soap bubble by top BK1 (blue curve) and bottom BK2 (red curve) microphones distant of 30~mm from the bubble center, averaged over five bursting events, the shaded areas being bounded by the maximum and minimum of the five signals. Dashed black curve: model accounting for acoustic emission by the retracting rim only (Eqs. ~\ref{eq:modeleforcelocalisee}, \ref{eq:force1}).  Solid black curve: model accounting for additional acoustic emission by the thickness shock wave propagating along the soap film (Eqs.~\ref{eq:modeleforcelocalisee}, \ref{eq:force2}).}
\label{fig2}
\end{figure}

\begin{figure*}[htbp]
\centering\includegraphics[width=170mm]{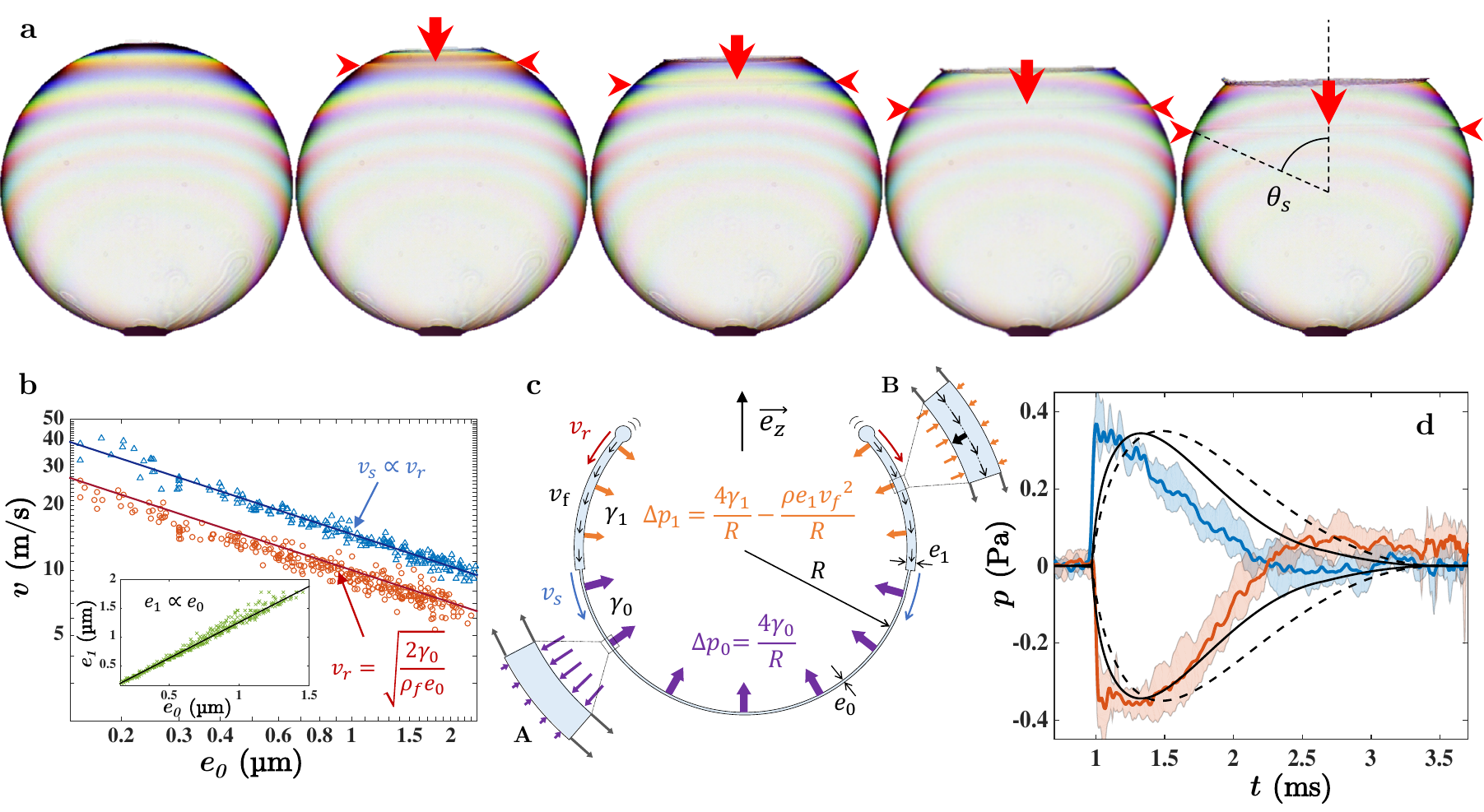}
\caption{\textbf{Observation of the thickness shock wave propagating along bursting long-lived bubbles and implications on the acoustic signature.} \textbf{a}, High-speed images of the spontaneous bursting of a long-lived, 0.5 mL soap bubble illuminated by a source of white light evidencing the propagation of a thickness shock wave downward ahead of the retracting rim, as indicated by red arrow tips. \textbf{b}, Variations of the thickness shock wave velocity $v_{\rm s}$ (blue symbols) and of the rim velocity $v_{\rm r}$ (red symbols) as function of the equilibrium film thickness $e_0$, evidencing that $v_{\rm s} \propto v_{\rm r}$ and demonstrating the validity of equation~(\ref{eq:vTC}). Inset: variation of soap film thickness between the shock wave and the rim $e_1(\theta,t)$ as function of the equilibrium film thickness $e_0(\theta)$, evidencing that $e_1$ is independent of time and $e_1 \propto e_0$. \textbf{c}, Sketch of the bursting soap bubble showing the rim retracting at velocity $v_{\rm r}$ (red arrow), the thickness shock wave propagating at velocity $v_{\rm s}$ (blue arrow) downward ahead of the rim and the thickened film flowing at velocity $v_{\rm f}$ (black arrows) behind the shock wave. The capillary stresses with amplitude $\Delta P_0$ (respectively $\Delta P_1$) exerted on the air by the film at rest (resp. by the flowing film) with thickness $e_0$ (resp. $e_1$) are represented as violet (resp. orange) arrows. Enlargement A: portion of the film at rest and at equilibrium showing the different pressure stresses (violet arrows) exerted by inner and outer air. Enlargement B: portion of the out-of-equilibrium, moving film showing the flow velocity (black arrows), the centripetal acceleration (bold, black arrow), the unchanged outer pressure stresses and the reduced inner pressure stresses (orange arrows). \textbf{d}, pressure signals acquired during the bursting of 0.5 mL, long-lived soap bubbles by top BK1 (blue curve) and bottom BK2 (red curve) microphones distant of 30~mm from the bubble center, averaged over three bursting events, the shaded areas being bounded by the maximum and minimum of the three signals. Dashed black curve: model accounting for acoustic emission by the retracting rim only (Eqs. ~\ref{eq:modeleforcelocalisee}, \ref{eq:force1}).  Solid black curve: model accounting for additional acoustic emission by the thickness shock wave propagating along the soap film. (Eqs.~\ref{eq:modeleforcelocalisee}, \ref{eq:force2}).}
\label{fig3}
\end{figure*}

We set out by blowing air bubbles with a syringe pump using a 0.25 g/L sodium dodecyl sulfate (SDS)-water solution. The bubbles of 1~mL typical volume are placed atop a vertical capillary tube (see Fig.~\ref{fig1}a).
Bursting events are either spontaneous or triggered using a hydrophobic needle located $10-15$~mm above the tube outlet, visible on top of the pictures of bursting bubble shown in Fig.~\ref{fig1}~a. Air inside the bubble is known to be pressurized at Laplace overpressure $\Delta P_0 = 4 \gamma_0/R$ by the two liquid-air interfaces constituting the soap film, $R \simeq 6$~mm being the bubble radius and $\gamma_0 = (50 \pm 1)~\text{mN} \cdot \text{m}^{-1}$ the equilibrium surface tension of the soap solution at ambient temperature. Therefore, one would expect an acoustic emission with spherical symmetry resulting from the sudden overpressure release following the film bursting and a N-shaped pressure signal with duration $2R/c \simeq 35~ \mu \rm s$ ($c \simeq 340~{\rm m} \cdot {\rm s}^{-1}$ is the speed of sound in air) similar to the popping sound of rubber balloons~\cite{Deihl1968} or to the blast wave of spherical explosions~\cite{Whitham1999}. To test the validity of this picture, we image the bubble bursting using a high-speed camera and we record simultaneously its acoustic emission using arrays of microphones, which allow us to determine the associated radiation pattern. In a first series of experiments dedicated to the identification of the sources of sound we use three circular acoustic antennae each made of eight calibrated MEMS microphones (see Sup. Mat.) regularly distributed along a planar, circular frame coinciding with a meridian of the bubble and with radius 44~mm, 71~mm and 112~mm respectively, sketched in Fig.~\ref{fig1}~b. The eight pressure signals recorded by the 44~mm radius antenna during the triggered bursting of a fresh bubble are shown in Fig.~\ref{fig1}~c. Remarkably, although the signals exhibit the same shape, their sign and amplitude depend on MEMS colatitude $\theta$ (as defined in Fig.~\ref{fig1}~b), the signals recorded above and below the bubble having opposite signs. Morevoer, we note that acoustic emission lasts for $4$~ms.
The puzzling disagreement between our observations and the naive picture sketched previously of isotropic acoustic radiation by the bursting bubble can be resolved by focusing on the pictures of the bubble during the bursting event. As known since Bull\cite{Bull1904}, bubble bursting usually begins with the spontaneous or triggered opening of a hole in the soap film followed by the growth of a circular rim gobbling up the soap film in its path at a typical speed $v_{\text{r}} \sim 10~\rm{m} \cdot {s}^{-1}$ up to its complete disappearance, as shown in the picture strip shown in Fig.~\ref{fig1}~a. According to \cite{Taylor1959a,Culick1960,Mcentee1969}, as the film is pulled by surface tension and its acceleration is moderated by the inertia of the growing liquid rim, $v_{\text{r}}$ is quantitatively predicted by \cite{Mcentee1969}:
\begin{equation}\label{eq:vTC}
    v_{\text{r}} = \sqrt{\frac{2 \gamma_0}{\rho_{\text{f}} e_0}},
\end{equation}
where $\rho_{\text{f}} = 1.0 \times 10^3~\text{kg} \cdot \text{m}^{-3}$ is the liquid film density and $e_0 \sim 1~\mu \text{m}$ its typical thickness. The film retraction is expected to last for $T = \pi R/v_{\text{r}} \simeq 5$~ms, in agreement with observations. The comparison between the pictures of the bursting bubble and the synchronized recording of its acoustic signature, both displayed in Fig.~\ref{fig1}~a and c, reveals that the bubble radiates sound during  whole opening. This suggests that acoustic emission may be governed by 
liquid film retraction. Actually, the physical mechanism of sound radiation can be elucidated by refering to aeroacoustics theory, which classifies sound emission into three kinds of sources, monopolar sources associated to the injection of mass into the air, dipolar sources due to momentum injection, and quadrupolar sources mainly associated to air flow \cite{Morse1968,Lighthill2001,Howe2014}. 
Here, during the bubble opening, the sudden relaxation of the air compressed in the bubble is expected to result in a monopolar acoustic emission (i.e. with spherical symmetry) of duration $T$ and pressure magnitude $p_M \sim \gamma_0 M^2/r$, where $M=v_{\text{r}}/c$ is the Mach number associated to the rim motion and $r$ the distance between the bubble center and the detection point (see Sup. Mat.). Moreover, throughout the soap film retraction the  stresses exerted by the soap film on the inner air $\Delta P_0 \, \bf{n}$ ($\bf{n}$ is the unit vector normal to the film oriented toward the bubble center) do not balance, as illustrated in the enlargement $A$ in Fig. \ref{fig3}~c. Their addition results in a capillary force $F \bf{e}_z$ directed upward ($\bf{e}_z$ is the vertical unit vector oriented upward). As the distribution of capillary stresses is unsteady, so is the resulting capillary force accelerating the air 
and an acoustic emission pointing along the direction of momentum injection occurs in the form of a dipolar radiation of duration $T$ with symmetry axis coinciding with the bubble center to initial hole axis, here the vertical. Given that here sound is recorded in the near-field of the acoustic source, i.e. at $r \lesssim cT$, the acoustic pressure amplitude associated to the dipolar radiation is $p_D \sim \gamma_0 R/r^2 \cos{\theta}$ (see Sup. Mat.). Finally, the air flow in the wake of the moving liquid rim should result in an acoustic emission of duration $T$ with quadrupolar symmetry and acoustic pressure amplitude in the near-field $p_Q \sim \frac{R^2}{r^3} e_0 \rho_{\text{a}} v_{\text{r}}^2$, where $\rho_{\text{a}} = 1.2~\text{kg}\cdot \text{m}^{-3}$ is air density (see Sup. Mat.). Given the investigated values of $r$ and $R$, $p_M/p_D \sim \frac{r}{R} \frac{4}{3 \pi^2} M^2 \sim 10^{-2}$ and $p_Q/p_D \sim \frac{R}{r} \frac{\rho_a}{\rho_e}  \sim 10^{-4}$, which explains why the measured acoustic signature of a bursting bubble presented in Fig.~\ref{fig1}~c appears as dipolar (conversely, this explains why bursting rubber balloons, which are actually teared along cracks commonly propagating at supersonic speeds even at moderate strains\cite{moulinet2015}, have acoustic signatures that appear as monopolar\cite{Deihl1968}). This conclusion can be strengthened by computing the three pressure fields associated respectively to the monopolar, dipolar and quadrupolar radiations from the 24 signals acquired using the acoustic antennae using a signal processing detailed in Sup. Mat. The corresponding three multipolar contributions to the pressure signal measured by the MEMS microphone located on the bubble top are shown in Fig.~\ref{fig1}~d. Their comparison confirms that the bubble acoustic emission is mainly dipolar, which leads us to focus on the dipolar radiation. According to Lighthill\cite{Lighthill2001}, assuming the bubble to disappear in the place, i.e. the film to remain spherical during bursting, away from the bubble, i.e. in the $r/R \gg 1$ limit, the acoustic pressure detected at distance $r$, colatitude $\theta$ and time laps $t$ after the onset of bubble bursting can be approximated by:
\begin{equation}\label{eq:modeleforcelocalisee}
    p_D(r,\theta,t) = \frac{1}{4\pi r} \cos{(\theta)} \left[ \frac{1}{c} \dot{F}(t^{'}) + \frac{1}{r} F(t^{'}) \right]
\end{equation}
with: 
\begin{equation}\label{eq:force1}
    F(t) = \pi R^2 \, \Delta P_0 \,   \sin^2{[\theta_{\text{r}}(t)]}
\end{equation}
$\theta_{\text{r}}(t)$ being the time-dependent colatitude of the retracting liquid rim, as defined in Fig.~\ref{fig2}~a, $t^{'}=t-r/c$ the retarded time and $\dot{F}$ the derivative of $F$ (see Sup. Mat.). In the frame of this simple model of bubble disappearance in the place by rim retraction, the capillary force varies only because the retracting rim continuously reduces the film surface area. Correlatively, the source of sound coincides with the rim. 

To quantitative test the validity of this model, in a second series of experiment we use two large bandwidth microphones (see Sup. Mat.) distant of $r=30$~mm from a $R=6.2$~mm bubble, i.e. in its near-field, and positionned above and below the bubble, named BK1 and BK2 in Fig.~\ref{fig1}. 
The two acoustic pressure signals acquired during the triggered burst of a fresh bubble are compared to their prediction using Eqs. ~\ref{eq:modeleforcelocalisee} and \ref{eq:force1} in Fig.~\ref{fig2}~b. $\theta_{\text{r}}(t)$ has been first extracted from the images of bubble bursting shown in Fig.~\ref{fig2}~a and then interpolated at acoustic sampling frequency. The model catches the shape and amplitude of the measured signals but overestimates their duration and maximal amplitude by approximatively $30~\%$ and $10~\%$ respectively. As shown in the following, the limitation of this model does not rely on the description of the acoustic emission process but actually on the roughness of the description of the film dynamics adopted up to now.

To describe the film dynamics more accurately, a time-resolved mapping of its thickness distribution during bursting is required. 
To this aim, long-lived bubbles are considered. Interestingly, due to gravity-driven drainage, long-lived bubbles indeed exhibit strong vertical thickness stratification and film thinning down to thicknesses comparable to optical wavelengths. Consequently, when illuminated by a source of white light and observed in transmission, long-lived bubbles display light interference fringes (see Fig.~\ref{fig3}~a) from which the film thickness distribution can be determined just before and during bursting (see Sup. Mat.). 

After typically one minute of lifetime, long-lived bubbles ultimately spontaneously burst by opening at their top where the film is the thinnest (see Fig.~\ref{fig3}~a). Consequently, the two microphones are always located along their axis of symmetry, as in needle-triggered bursting experiments. The detected acoustic signals, shown in Fig.~\ref{fig3}~d, are initially much steeper than in the case of fresh bubbles. This may be ascribed to the initially large retraction velocity $v_{\text{r}} \propto e_0^{-1/2}$ (see equation~\ref{eq:vTC}) of the thin film located on the bubble top combined with the dependence of the radiated sound on $\dot{F} \propto \dot{\theta_{\text{r}}} = v_{\text{r}}/R$ (see equation~\ref{eq:modeleforcelocalisee}). Once the actual rim colatitude evolution $\theta_{\rm r}(t)$ determined by image analysis and interpolated, the model based on Eqs. ~\ref{eq:modeleforcelocalisee}, \ref{eq:force1} can be used to predict these acoustic signals. As shown in Fig.~\ref{fig3}~d, the agreement between experimental data and their prediction is qualitative, although satisfactory, as in the case of fresh bubbles. 

A careful observation of the light interference pattern displayed by long-lived bursting bubbles during bursting reveals that, starting from the hole, a thickness discontinuity propagates along the film downward ahead of the rim, as indicated by red arrow tips in the pictures shown in Fig.~\ref{fig3}~a. The generation of such a thickness shock wave during the retraction of soap films is commonly observed, in particular along planar films made of the same soap solution \cite{Mcentee1969}. The decrease of the soap film area resulting from the rim retraction indeed induces a rapid compression of the surfactants located at the film surface. Since their desorption kinetics is generally slow compared to the film retraction velocity, surfactants behave as if they were insoluble \cite{Vijayendran1975}. Consequently they are carried along the compressed films and concentrate. Due to this rapid surfactant concentration, the surface tension of the compressed films decreases and departs from its equilibrium value. Depending on the out-of-equilibrium surface tension versus surface compression relation specific to the surfactant, a thickness and surface tension shock wave may appear \cite{Frankel1969}, as sketched in Fig.~\ref{fig3}~c, which is the case for SDS in our experiments and in \cite{Mcentee1969}. 

The occurrence of a thickness shock wave during bubble bursting experiments is expected to affect acoustic emission in several aspects. First, recalling that acoustic emission originates from the unsteadiness of the capillary stresses, we identify this surface tension discontinuity propagating in front of the rim as a supplementary source of sound. More specifically, the surface tension drop behind the thickness shock wave tends to increase the capillary force exerted by the bubble on the inner air (and hence the acoustic emission) when the shock wave is located in the upper hemisphere ($\theta < \pi/2$), as sketched in Fig.~\ref{fig3}~c, and to decrease it in the lower hemisphere ($\theta > \pi/2$), in agreement with the observed evolution of the sign of the difference between the experimental signals and their prediction based on Eqs. ~\ref{eq:modeleforcelocalisee}, \ref{eq:force1} shown in Fig.~\ref{fig3}~d.

Moreover, the shock wave triggers a flow of the soap film that further decreases the capillary stresses, as demonstrated in the following. Assuming a thickness shock wave located at colatitude $\theta_{\text{s}}(t)$ to propagate at velocity $v_{\text{s}}$ along a film initially at rest and at equilibrium, surfactant effective insolubility assumption, mass conservation and momentum balance across the shock indeed entail the motion of the whole film behind the shock wave at velocity $v_{\text{f}}(\theta_{\text{s}}^-)$ that satisfies the following relations \cite{Frankel1969}:
\begin{eqnarray}
v_{\text{f}}(\theta_{\text{s}}^-) &=& v_{\text{s}} \left[ 1-\frac{e_0(\theta_{\text{s}}^+)}{e_1(\theta_{\text{s}}^{-})} \right], \label{eq:masscons}\\
2 \left[ \gamma_0(\theta_{\text{s}}^+) -\gamma_1(\theta_{\text{s}}^{-}) \right] &=& \rho_{\text{f}} \; e_0(\theta_{\text{s}}^+) \;  v_{\text{f}}(\theta_{\text{s}}^-) \;  v_{\text{s}},\label{eq:momentumcons}
\end{eqnarray}
where $e_0(\theta_{\text{s}}^+)$ is the thickness of the film at rest and at equilibrium in front of the shock wave, $e_1(\theta_{\text{s}}^{-})$ and $\gamma_1(\theta_{\text{s}}^{-})$ the thickness of the compressed film and its out-of-equilibrium surface tension behind the shock wave, respectively. Consequently, behind the shock, each moving fluid element runs a circular trajectory with radius $R$ around the bubble center at tangential velocity $v_{\text{f}}$ and is thus subjected to the centripetal acceleration $v_{\text{f}}^2/R$. Application of second Newton's law to such an accelerated fluid element with thickness $e_1$ submitted to external pressure $P_0$, internal pressure $P_{\text{in}}$ and surface tension $\gamma_1$, as sketched in the enlargement B in Fig.~\ref{fig3}~c, reveals that the centripetal acceleration of the moving film noticeably reduces the overpressure it exerts on the inner air, i.e. $P_{\text{in}}-P_0 = \Delta P_1 = 4 \, \gamma_1/R - \rho_{\text{f}} e_1 v_{\text{f}}^2/R < 4 \, \gamma_1/R$. Such an influence of the in-plane film motion on the pressure jump across curved films is known to be responsible of the large variety of shapes of water bells \cite{Savart1833b,Savart1833c,Boussinesq1869,Taylor1959}.


All the quantitites involved in this description of the soap film dynamics can be experimentally assessed by analysing the light interference patterns displayed by the bubble. First, the rim colatitude $\theta_{\text{r}}(t)$, the shock wave colatitude $\theta_{\text{s}}(t)$, the film thickness distribution $e_0(\theta)$ along the film at equilibrium and at rest 
and the film thickness distribution $e_1(\theta,t)$ along the out-of-equilibrium, flowing film behind the shock are experimentally determined from image analysis. Next, neglecting the weak departure to sphericity of the film during its retraction, the rim and shock wave velocities are calculated using $v_{\text{r}} = R \dot{\theta}_{\text{r}}$ and $v_{\text{s}} = R \dot{\theta}_{\text{s}}$. Finally, $v_{\text{f}}(\theta_{\text{s}}^-)$ and $\gamma_1(\theta_{\text{s}}^-)$ are calculated using Eqs.~\ref{eq:masscons}, \ref{eq:momentumcons}. The agreement between the experimental variation of $v_{\text{r}}(\theta_{\text{r}})$ with $e_0(\theta_{\text{r}})$ and its prediction using equation~(\ref{eq:vTC}), both shown in Fig.~\ref{fig3}~b, demonstrates the validity of equation~(\ref{eq:vTC}) along a curved and stratified soap film.
As also shown in Fig.~\ref{fig3}~b, $v_{\text{s}} \simeq \alpha \, v_{\text{r}}$ with $\alpha = 1.57$, in quantitative agreement with previous measurements performed along planar soap films~\cite{Mcentee1969}. Next, the dependence of $e_1(\theta,t)$ versus $e_0(\theta)$ all along the moving film, plotted in the inset of Fig.~\ref{fig3}, is found to be linear and exhibits little scattering, indicating that $e_1$ does not basically depend on time and satisfies $e_1 (\theta) \simeq \beta \, e_0(\theta)$ with $\beta = 1.27$. As detailled in Supp. Mat., from these experimental results it can be shown that, given the vality of equation~(\ref{eq:vTC}) and since $v_{\text{s}}/v_{\text{r}}$ and $e_1/e_0$ are constant along the moving film, $v_{\text{f}}(\theta_{\text{s}}^-)$ and $\gamma_1(\theta_{\text{s}}^{-})$ can be determined from the sole measurement of the rim retraction kinetics $\theta_{\text{r}}(t)$ and the independent measurement of $\gamma_0$ using Eqs.~\ref{eq:masscons}, \ref{eq:momentumcons}. Finally, as detailled in Sup. Mat., surfactant insolubility assumption allows one to show that $\Delta P_1$ is constant all along the moving film. This leads us to conclude that the only sources of unsteadiness of the capillary stresses, and therefore of acoustic emission, are the moving rim and shock wave. The capillary stresses along the bursting bubble being constant equal to $\Delta P_1$ in the range $\theta \in [\theta_{\text{r}}(t) ; \theta_{\text{s}}(t)]$ and to $\Delta P_0$ in the range $\theta \in [\theta_{\text{s}}(t) ; \pi]$, their addition results in the following capillary force exerted by the soap film on the inner air: 
\begin{equation}
\label{eq:force2}
F(t)=\pi R^2 \left\{ (\Delta P_0 - \Delta P_1) \sin^2{[\theta_{\text{s}}(t)]} + \Delta P_1 \sin^2{[\theta_{\text{r}}(t)]} \right \}
\end{equation}
from wich the acoustic emission away from the bubble can be evaluated using equation~(\ref{eq:modeleforcelocalisee}). The agreement between the acoustic signals detected by BK1 and BK2 microphones during the bursting of long-lived bubble and their prediction using equation~(\ref{eq:force2}), shown in Fig.~\ref{fig3}~d, is more satisfactory than using equation~(\ref{eq:force1}), in particular regarding the signal duration. 

Extrapolating this refined description of film dynamics to fresh bubbles allows one to predict and to quantitatively describe the propagation of a thickness shock wave along bursting fresh bubbles, although they are made of thick films that display no light interference fringes and hardly visible shock waves. This is possible because the whole description of the film dynamics requires only the knowledge of the rim colatitude $\theta_{\text{r}}(t)$ that is as easily measureable on fresh bubbles as on long-lived ones. When applying this refined model of acoutic emission (Eqs.~\ref{eq:modeleforcelocalisee}, \ref{eq:force2}) to fresh bubbles, one observes a quantitative agreement between the acoustic signals measured during the needdle-triggered bursting of fresh bubbles and their prediction, as shown in Fig.~\ref{fig2}~b. This confirms that the thickness shock wave propagating along the bursting soap film noticeably contributes to sound emission in the case of fresh bubbles too. Thanks to this thorough study, we have shown that the description of the acoustic emission by a bursting bubble we propose is fully consistent with the current understanding of soap film retraction dynamics.

This study has revealed the profusion of information on bubble bursting carried by the sound radiated during the event and accessible using aeroacoustics theory. In particular, we could reveal the acoustic signature of the thickness shock wave propagating along the bursting soap film. As exemplified by this study of a specific hydrodynamic event, dipolar acoustic radiation can inform us about the forces at play during the rapid evolution of liquid interfaces and more generally during violent events, thus potentially constituting a precious diagnostic complementary to high-speed imaging that primarily reveals shapes. However, such an acoustic radiation constitutes only a tiny record of the violence of the bubble bursting event and not a mechanism of energy loss, since the ratio of the total acoustic energy radiated to the variation of surface energy\cite{Pandit1990,deGennes1996} is smaller than $10^{-6}$, see Sup. Mat. for details. 

\textbf{Methods}

\footnotesize{

\textbf{Soap bubble production.} Sodium dodecyl sulfate (SDS) from {\it Euromedex} was used as surfactant and was diluted in distilled water down to 0.25~g/L. The surface tension of the solution was measured using the du No\"{u}y ring method. 1.5~$\mu$L droplets of the soap solution were carefully deposited on top of a vertical capillary tube using a micropipette. To limit the draining of the solution before bubble inflation, the end of the tube was thickened using a rim of glue. Bubbles were then inflated at a calibrated volume using a seringe pump with a rate of 20~mL/min. For triggering the bursting of fresh bubbles, a hydrophobic Rain X\textsuperscript{\textregistered}-coated needle was positionned vertically at a distance of 2$R$ above the capillary end.

\textbf{Video and sound recordings.} High-speed imaging of bursting bubbles was performed using a {\it Photron Fastcam SA5} camera at a 20 000 fps rate when color movies were required and a {\it Vision Research Phantom V711} camera at a 25 000 fps rate for the black and white movies. Video and sound recordings were synchronized by triggering the acquisition of the acoustic signals once the bubble top disappeared on the bubble pictures using the Image Triggering function of the camera software.
}

\bibliographystyle{apsrev4-1}
\bibliography{Biblio}

\begin{thebibliography}{31}%
\makeatletter
\providecommand \@ifxundefined [1]{%
 \@ifx{#1\undefined}
}%
\providecommand \@ifnum [1]{%
 \ifnum #1\expandafter \@firstoftwo
 \else \expandafter \@secondoftwo
 \fi
}%
\providecommand \@ifx [1]{%
 \ifx #1\expandafter \@firstoftwo
 \else \expandafter \@secondoftwo
 \fi
}%
\providecommand \natexlab [1]{#1}%
\providecommand \enquote  [1]{``#1''}%
\providecommand \bibnamefont  [1]{#1}%
\providecommand \bibfnamefont [1]{#1}%
\providecommand \citenamefont [1]{#1}%
\providecommand \href@noop [0]{\@secondoftwo}%
\providecommand \href [0]{\begingroup \@sanitize@url \@href}%
\providecommand \@href[1]{\@@startlink{#1}\@@href}%
\providecommand \@@href[1]{\endgroup#1\@@endlink}%
\providecommand \@sanitize@url [0]{\catcode `\\12\catcode `\$12\catcode
  `\&12\catcode `\#12\catcode `\^12\catcode `\_12\catcode `\%12\relax}%
\providecommand \@@startlink[1]{}%
\providecommand \@@endlink[0]{}%
\providecommand \url  [0]{\begingroup\@sanitize@url \@url }%
\providecommand \@url [1]{\endgroup\@href {#1}{\urlprefix }}%
\providecommand \urlprefix  [0]{URL }%
\providecommand \Eprint [0]{\href }%
\providecommand \doibase [0]{http://dx.doi.org/}%
\providecommand \selectlanguage [0]{\@gobble}%
\providecommand \bibinfo  [0]{\@secondoftwo}%
\providecommand \bibfield  [0]{\@secondoftwo}%
\providecommand \translation [1]{[#1]}%
\providecommand \BibitemOpen [0]{}%
\providecommand \bibitemStop [0]{}%
\providecommand \bibitemNoStop [0]{.\EOS\space}%
\providecommand \EOS [0]{\spacefactor3000\relax}%
\providecommand \BibitemShut  [1]{\csname bibitem#1\endcsname}%
\let\auto@bib@innerbib\@empty
\bibitem [{\citenamefont {Houle}\ and\ \citenamefont
  {Sethna}(1996)}]{Houle1996}%
  \BibitemOpen
  \bibfield  {author} {\bibinfo {author} {\bibfnamefont {P.~A.}\ \bibnamefont
  {Houle}}\ and\ \bibinfo {author} {\bibfnamefont {J.~P.}\ \bibnamefont
  {Sethna}},\ }\href {\doibase 10.1103/PhysRevE.54.278} {\bibfield  {journal}
  {\bibinfo  {journal} {Phys. Rev. E}\ }\textbf {\bibinfo {volume} {54}},\
  \bibinfo {pages} {278} (\bibinfo {year} {1996})}\BibitemShut {NoStop}%
\bibitem [{\citenamefont {Salminen}\ \emph {et~al.}(2002)\citenamefont
  {Salminen}, \citenamefont {Tolvanen},\ and\ \citenamefont
  {Alava}}]{Salminen2002}%
  \BibitemOpen
  \bibfield  {author} {\bibinfo {author} {\bibfnamefont {L.~I.}\ \bibnamefont
  {Salminen}}, \bibinfo {author} {\bibfnamefont {A.~I.}\ \bibnamefont
  {Tolvanen}}, \ and\ \bibinfo {author} {\bibfnamefont {M.~J.}\ \bibnamefont
  {Alava}},\ }\href {\doibase 10.1103/PhysRevLett.89.185503} {\bibfield
  {journal} {\bibinfo  {journal} {Phys. Rev. Lett.}\ }\textbf {\bibinfo
  {volume} {89}},\ \bibinfo {pages} {185503} (\bibinfo {year}
  {2002})}\BibitemShut {NoStop}%
\bibitem [{\citenamefont {Rubinstein}\ \emph {et~al.}(2004)\citenamefont
  {Rubinstein}, \citenamefont {Cohen},\ and\ \citenamefont
  {Fineberg}}]{Rubinstein2004}%
  \BibitemOpen
  \bibfield  {author} {\bibinfo {author} {\bibfnamefont {S.~M.}\ \bibnamefont
  {Rubinstein}}, \bibinfo {author} {\bibfnamefont {G.}~\bibnamefont {Cohen}}, \
  and\ \bibinfo {author} {\bibfnamefont {J.}~\bibnamefont {Fineberg}},\ }\href
  {\doibase 10.1038/nature02830} {\bibfield  {journal} {\bibinfo  {journal}
  {Nature}\ }\textbf {\bibinfo {volume} {430}},\ \bibinfo {pages} {1005}
  (\bibinfo {year} {2004})}\BibitemShut {NoStop}%
\bibitem [{\citenamefont {Franz}(1959)}]{Franz1959}%
  \BibitemOpen
  \bibfield  {author} {\bibinfo {author} {\bibfnamefont {G.~J.}\ \bibnamefont
  {Franz}},\ }\href {\doibase 10.1121/1.1907831} {\bibfield  {journal}
  {\bibinfo  {journal} {J. Acoust. Soc. Am.}\ }\textbf {\bibinfo {volume}
  {31}},\ \bibinfo {pages} {1080} (\bibinfo {year} {1959})}\BibitemShut
  {NoStop}%
\bibitem [{\citenamefont {Gordillo}\ \emph {et~al.}(2018)\citenamefont
  {Gordillo}, \citenamefont {Sun},\ and\ \citenamefont {Cheng}}]{Gordillo2018}%
  \BibitemOpen
  \bibfield  {author} {\bibinfo {author} {\bibfnamefont {L.}~\bibnamefont
  {Gordillo}}, \bibinfo {author} {\bibfnamefont {T.-P.}\ \bibnamefont {Sun}}, \
  and\ \bibinfo {author} {\bibfnamefont {X.}~\bibnamefont {Cheng}},\ }\href
  {\doibase 10.1017/jfm.2017.901} {\bibfield  {journal} {\bibinfo  {journal}
  {J. Fluid Mech.}\ }\textbf {\bibinfo {volume} {840}},\ \bibinfo {pages} {190}
  (\bibinfo {year} {2018})}\BibitemShut {NoStop}%
\bibitem [{\citenamefont {Aljishi}\ and\ \citenamefont
  {Tatarkiewicz}(1991)}]{Aljishi1991}%
  \BibitemOpen
  \bibfield  {author} {\bibinfo {author} {\bibfnamefont {S.}~\bibnamefont
  {Aljishi}}\ and\ \bibinfo {author} {\bibfnamefont {J.}~\bibnamefont
  {Tatarkiewicz}},\ }\href {\doibase 10.1119/1.16784} {\bibfield  {journal}
  {\bibinfo  {journal} {Am. J. Phys.}\ }\textbf {\bibinfo {volume} {59}},\
  \bibinfo {pages} {628} (\bibinfo {year} {1991})}\BibitemShut {NoStop}%
\bibitem [{\citenamefont {Rayleigh}(1891)}]{Rayleigh1891}%
  \BibitemOpen
  \bibfield  {author} {\bibinfo {author} {\bibfnamefont {L.}~\bibnamefont
  {Rayleigh}},\ }\href {\doibase 10.1038/044249e0} {\bibfield  {journal}
  {\bibinfo  {journal} {Nature}\ }\textbf {\bibinfo {volume} {44}},\ \bibinfo
  {pages} {249} (\bibinfo {year} {1891})}\BibitemShut {NoStop}%
\bibitem [{\citenamefont {Savart}(1833{\natexlab{a}})}]{Savart1833}%
  \BibitemOpen
  \bibfield  {author} {\bibinfo {author} {\bibfnamefont {F.}~\bibnamefont
  {Savart}},\ }\href@noop {} {\bibfield  {journal} {\bibinfo  {journal} {Ann.
  Chim. Phys.}\ }\textbf {\bibinfo {volume} {53}},\ \bibinfo {pages} {337}
  (\bibinfo {year} {1833}{\natexlab{a}})}\BibitemShut {NoStop}%
\bibitem [{\citenamefont {Worthington}(1908)}]{Worthington1908}%
  \BibitemOpen
  \bibfield  {author} {\bibinfo {author} {\bibfnamefont {A.}~\bibnamefont
  {Worthington}},\ }\href@noop {} {\emph {\bibinfo {title} {A study of
  splashes}}}\ (\bibinfo  {publisher} {Longmans, Green and co.},\ \bibinfo
  {year} {1908})\BibitemShut {NoStop}%
\bibitem [{\citenamefont {Lefebvre}\ and\ \citenamefont
  {Mannoni}(1995)}]{Lefebvre1995}%
  \BibitemOpen
  \bibfield  {author} {\bibinfo {author} {\bibfnamefont {T.}~\bibnamefont
  {Lefebvre}}\ and\ \bibinfo {author} {\bibfnamefont {L.}~\bibnamefont
  {Mannoni}},\ }\href {\doibase 10.3406/1895.1995.1112} {\bibfield  {journal}
  {\bibinfo  {journal} {1895 Mille huit cent quatre-vingt-quinze}\ }\textbf
  {\bibinfo {volume} {18}},\ \bibinfo {pages} {144} (\bibinfo {year}
  {1995})}\BibitemShut {NoStop}%
\bibitem [{\citenamefont {Engel}(1955)}]{Engel1955}%
  \BibitemOpen
  \bibfield  {author} {\bibinfo {author} {\bibfnamefont {O.~G.}\ \bibnamefont
  {Engel}},\ }\href@noop {} {\bibfield  {journal} {\bibinfo  {journal} {J. Res.
  Nat. Bur. Stand.}\ }\textbf {\bibinfo {volume} {54}},\ \bibinfo {pages} {281}
  (\bibinfo {year} {1955})}\BibitemShut {NoStop}%
\bibitem [{\citenamefont {Ranz}(1959)}]{Ranz1959}%
  \BibitemOpen
  \bibfield  {author} {\bibinfo {author} {\bibfnamefont {W.~E.}\ \bibnamefont
  {Ranz}},\ }\href {\doibase 10.1063/1.1735095} {\bibfield  {journal} {\bibinfo
   {journal} {J. Appl. Phys.}\ }\textbf {\bibinfo {volume} {30}},\ \bibinfo
  {pages} {1950} (\bibinfo {year} {1959})}\BibitemShut {NoStop}%
\bibitem [{\citenamefont {Thoroddsen}\ \emph {et~al.}(2008)\citenamefont
  {Thoroddsen}, \citenamefont {Etoh},\ and\ \citenamefont
  {Takehara}}]{Thoroddsen2008}%
  \BibitemOpen
  \bibfield  {author} {\bibinfo {author} {\bibfnamefont {S.}~\bibnamefont
  {Thoroddsen}}, \bibinfo {author} {\bibfnamefont {T.}~\bibnamefont {Etoh}}, \
  and\ \bibinfo {author} {\bibfnamefont {K.}~\bibnamefont {Takehara}},\ }\href
  {\doibase 10.1146/annurev.fluid.40.111406.102215} {\bibfield  {journal}
  {\bibinfo  {journal} {Ann. Rev. Fluid Mech.}\ }\textbf {\bibinfo {volume}
  {40}},\ \bibinfo {pages} {257} (\bibinfo {year} {2008})}\BibitemShut
  {NoStop}%
\bibitem [{\citenamefont {Deihl}\ and\ \citenamefont
  {F.~Roy~Carlson}(1968)}]{Deihl1968}%
  \BibitemOpen
  \bibfield  {author} {\bibinfo {author} {\bibfnamefont {D.~T.}\ \bibnamefont
  {Deihl}}\ and\ \bibinfo {author} {\bibfnamefont {J.}~\bibnamefont
  {F.~Roy~Carlson}},\ }\href {\doibase 10.1119/1.1974556} {\bibfield  {journal}
  {\bibinfo  {journal} {Am. J. Phys.}\ }\textbf {\bibinfo {volume} {36}},\
  \bibinfo {pages} {441} (\bibinfo {year} {1968})}\BibitemShut {NoStop}%
\bibitem [{\citenamefont {Whitham}(1999)}]{Whitham1999}%
  \BibitemOpen
  \bibfield  {author} {\bibinfo {author} {\bibfnamefont {G.~B.}\ \bibnamefont
  {Whitham}},\ }\href@noop {} {\emph {\bibinfo {title} {Linear and nonlinear
  waves}}},\ Vol.~\bibinfo {volume} {42}\ (\bibinfo  {publisher} {John Wiley \&
  Sons},\ \bibinfo {year} {1999})\BibitemShut {NoStop}%
\bibitem [{\citenamefont {Bull}(1904)}]{Bull1904}%
  \BibitemOpen
  \bibfield  {author} {\bibinfo {author} {\bibfnamefont {L.}~\bibnamefont
  {Bull}},\ }\href@noop {} {\bibfield  {journal} {\bibinfo  {journal} {Institut
  E.-J. Marey}\ } (\bibinfo {year} {1904})}\BibitemShut {NoStop}%
\bibitem [{\citenamefont {Taylor}(1959{\natexlab{a}})}]{Taylor1959a}%
  \BibitemOpen
  \bibfield  {author} {\bibinfo {author} {\bibfnamefont {G.~I.}\ \bibnamefont
  {Taylor}},\ }\href@noop {} {\bibfield  {journal} {\bibinfo  {journal} {Proc.
  R. Soc. Lond. A}\ }\textbf {\bibinfo {volume} {253}},\ \bibinfo {pages} {313}
  (\bibinfo {year} {1959}{\natexlab{a}})}\BibitemShut {NoStop}%
\bibitem [{\citenamefont {Culick}(1960)}]{Culick1960}%
  \BibitemOpen
  \bibfield  {author} {\bibinfo {author} {\bibfnamefont {F.~E.~C.}\
  \bibnamefont {Culick}},\ }\href@noop {} {\bibfield  {journal} {\bibinfo
  {journal} {J. AppL. Phys.}\ }\textbf {\bibinfo {volume} {31}},\ \bibinfo
  {pages} {1128} (\bibinfo {year} {1960})}\BibitemShut {NoStop}%
\bibitem [{\citenamefont {McEntee}\ and\ \citenamefont
  {Mysels}(1969)}]{Mcentee1969}%
  \BibitemOpen
  \bibfield  {author} {\bibinfo {author} {\bibfnamefont {W.~R.}\ \bibnamefont
  {McEntee}}\ and\ \bibinfo {author} {\bibfnamefont {K.~J.}\ \bibnamefont
  {Mysels}},\ }\href@noop {} {\bibfield  {journal} {\bibinfo  {journal} {J.
  Phys. Chem.}\ }\textbf {\bibinfo {volume} {73}},\ \bibinfo {pages} {3018}
  (\bibinfo {year} {1969})}\BibitemShut {NoStop}%
\bibitem [{\citenamefont {Morse}\ and\ \citenamefont
  {Ingard}(1968)}]{Morse1968}%
  \BibitemOpen
  \bibfield  {author} {\bibinfo {author} {\bibfnamefont {P.~M.}\ \bibnamefont
  {Morse}}\ and\ \bibinfo {author} {\bibfnamefont {K.~U.}\ \bibnamefont
  {Ingard}},\ }\href@noop {} {\emph {\bibinfo {title} {Theoretical
  acoustics}}}\ (\bibinfo  {publisher} {Princeton university press},\ \bibinfo
  {year} {1968})\BibitemShut {NoStop}%
\bibitem [{\citenamefont {Lighthill}(2001)}]{Lighthill2001}%
  \BibitemOpen
  \bibfield  {author} {\bibinfo {author} {\bibfnamefont {J.}~\bibnamefont
  {Lighthill}},\ }\href@noop {} {\emph {\bibinfo {title} {Waves in Fluids}}}\
  (\bibinfo  {publisher} {Cambridge University Press},\ \bibinfo {year}
  {2001})\BibitemShut {NoStop}%
\bibitem [{\citenamefont {Howe}(2014)}]{Howe2014}%
  \BibitemOpen
  \bibfield  {author} {\bibinfo {author} {\bibfnamefont {M.}~\bibnamefont
  {Howe}},\ }\href@noop {} {\emph {\bibinfo {title} {Acoustics and Aerodynamic
  Sound}}}\ (\bibinfo  {publisher} {Cambridge University Press},\ \bibinfo
  {year} {2014})\BibitemShut {NoStop}%
\bibitem [{\citenamefont {Moulinet}\ and\ \citenamefont
  {Adda-Bedia}(2015)}]{moulinet2015}%
  \BibitemOpen
  \bibfield  {author} {\bibinfo {author} {\bibfnamefont {S.}~\bibnamefont
  {Moulinet}}\ and\ \bibinfo {author} {\bibfnamefont {M.}~\bibnamefont
  {Adda-Bedia}},\ }\href@noop {} {\bibfield  {journal} {\bibinfo  {journal}
  {Phys. Rev. Lett.}\ }\textbf {\bibinfo {volume} {115}},\ \bibinfo {pages}
  {184301} (\bibinfo {year} {2015})}\BibitemShut {NoStop}%
\bibitem [{\citenamefont {Vijayendran}(1975)}]{Vijayendran1975}%
  \BibitemOpen
  \bibfield  {author} {\bibinfo {author} {\bibfnamefont {B.~R.}\ \bibnamefont
  {Vijayendran}},\ }\href@noop {} {\bibfield  {journal} {\bibinfo  {journal}
  {J. Phys. Chem.}\ }\textbf {\bibinfo {volume} {79}},\ \bibinfo {pages} {2501}
  (\bibinfo {year} {1975})}\BibitemShut {NoStop}%
\bibitem [{\citenamefont {Frankel}\ and\ \citenamefont
  {Mysels}(1969)}]{Frankel1969}%
  \BibitemOpen
  \bibfield  {author} {\bibinfo {author} {\bibfnamefont {S.}~\bibnamefont
  {Frankel}}\ and\ \bibinfo {author} {\bibfnamefont {K.~J.}\ \bibnamefont
  {Mysels}},\ }\href@noop {} {\bibfield  {journal} {\bibinfo  {journal} {J.
  Phys. Chem.}\ }\textbf {\bibinfo {volume} {73}},\ \bibinfo {pages} {3028}
  (\bibinfo {year} {1969})}\BibitemShut {NoStop}%
\bibitem [{\citenamefont {Savart}(1833{\natexlab{b}})}]{Savart1833b}%
  \BibitemOpen
  \bibfield  {author} {\bibinfo {author} {\bibfnamefont {F.}~\bibnamefont
  {Savart}},\ }\href@noop {} {\bibfield  {journal} {\bibinfo  {journal} {Ann.
  Chim}\ }\textbf {\bibinfo {volume} {54}},\ \bibinfo {pages} {56} (\bibinfo
  {year} {1833}{\natexlab{b}})}\BibitemShut {NoStop}%
\bibitem [{\citenamefont {Savart}(1833{\natexlab{c}})}]{Savart1833c}%
  \BibitemOpen
  \bibfield  {author} {\bibinfo {author} {\bibfnamefont {F.}~\bibnamefont
  {Savart}},\ }\href@noop {} {\bibfield  {journal} {\bibinfo  {journal} {Ann.
  Chim.}\ }\textbf {\bibinfo {volume} {54}},\ \bibinfo {pages} {113} (\bibinfo
  {year} {1833}{\natexlab{c}})}\BibitemShut {NoStop}%
\bibitem [{\citenamefont {Boussinesq}(1869)}]{Boussinesq1869}%
  \BibitemOpen
  \bibfield  {author} {\bibinfo {author} {\bibfnamefont {J.}~\bibnamefont
  {Boussinesq}},\ }\href@noop {} {\bibfield  {journal} {\bibinfo  {journal} {CR
  Acad. Sci. Paris}\ }\textbf {\bibinfo {volume} {69}},\ \bibinfo {pages} {45}
  (\bibinfo {year} {1869})}\BibitemShut {NoStop}%
\bibitem [{\citenamefont {Taylor}(1959{\natexlab{b}})}]{Taylor1959}%
  \BibitemOpen
  \bibfield  {author} {\bibinfo {author} {\bibfnamefont {G.~I.}\ \bibnamefont
  {Taylor}},\ }\href@noop {} {\bibfield  {journal} {\bibinfo  {journal} {Proc.
  R. Soc. Lond. A}\ }\textbf {\bibinfo {volume} {253}},\ \bibinfo {pages} {289}
  (\bibinfo {year} {1959}{\natexlab{b}})}\BibitemShut {NoStop}%
\bibitem [{\citenamefont {Pandit}\ and\ \citenamefont
  {Davidson}(1990)}]{Pandit1990}%
  \BibitemOpen
  \bibfield  {author} {\bibinfo {author} {\bibfnamefont {A.~B.}\ \bibnamefont
  {Pandit}}\ and\ \bibinfo {author} {\bibfnamefont {J.~F.}\ \bibnamefont
  {Davidson}},\ }\href {\doibase 10.1017/S0022112090001823} {\bibfield
  {journal} {\bibinfo  {journal} {J. Fluid Mech.}\ }\textbf {\bibinfo {volume}
  {212}},\ \bibinfo {pages} {11} (\bibinfo {year} {1990})}\BibitemShut
  {NoStop}%
\bibitem [{\citenamefont {de~Gennes}(1996)}]{deGennes1996}%
  \BibitemOpen
  \bibfield  {author} {\bibinfo {author} {\bibfnamefont {P.-G.}\ \bibnamefont
  {de~Gennes}},\ }\href@noop {} {\bibfield  {journal} {\bibinfo  {journal}
  {Faraday Discuss.}\ }\textbf {\bibinfo {volume} {104}},\ \bibinfo {pages} {1}
  (\bibinfo {year} {1996})}\BibitemShut {NoStop}%
\end{thebibliography}%

\end{document}